\documentclass[sigconf]{acmart}

\usepackage{booktabs} 
\usepackage{amsmath,bm}
\usepackage{mathrsfs}
\usepackage{multirow}
\usepackage{tabularx}
\usepackage{subfig}
\usepackage{bbm}
\usepackage[utf8]{inputenc}
\usepackage{hyperref}
\usepackage[textsize=scriptsize]{todonotes}

\def\BibTeX{{\rm B\kern-.05em{\sc i\kern-.025em b}\kern-.08emT\kern-.1667em\lower.7ex\hbox{E}\kern-.125emX}}
    
\begin{document}

%
\title{CSRN: Collaborative Sequential Recommendation Networks\\ for News Retrieval}

%
\author{Bing Bai$^1$, Guanhua Zhang$^{12}$, Ye Lin$^1$, Hao Li$^1$, Kun Bai$^1$, Bo Luo$^3$}
\affiliation{%
  \institution{$^1$Cloud and Smart Industries Group, Tencent, China}
  \institution{$^2$Harbin Institute of Technology, China}
  \institution{$^3$The University of Kansas, USA}
}
\email{{icebai, guanhzhang, yessicalin, leehaoli, kunbai}@tencent.com, bluo@ku.edu}

%
\renewcommand{\shortauthors}{Bai and Zhang, et al.}

%
\begin{abstract}


Nowadays, news apps have taken over the popularity of paper-based media, providing a great opportunity for personalization.
Recurrent~Neural~Network~(RNN)-based sequential recommendation is a popular approach that utilizes users' recent browsing history to predict future items. 
This approach is limited that it does not consider the societal influences of news consumption, i.e., users may follow popular topics that are constantly changing, while certain hot topics might be spreading only among specific groups of people. 
Such societal impact is difficult to predict given only users' own reading histories. On the other hand, the traditional User-based~Collaborative~Filtering~(UserCF) makes recommendations based on the interests of the ``neighbors'', which provides the possibility to supplement the weaknesses of RNN-based methods. However, conventional UserCF only uses a single similarity metric to model the relationships between users, which is too coarse-grained and thus limits the performance.
In this paper, we propose a framework of deep neural networks to integrate the RNN-based sequential recommendations and the key ideas from UserCF, to develop Collaborative~Sequential~Recommendation~Networks~(CSRNs). Firstly, we build a directed co-reading network of users, to capture the fine-grained topic-specific similarities between users in a vector space. Then, the CSRN model encodes users with RNNs, and learns to attend to neighbors and summarize what news they are reading at the moment. Finally, news articles are recommended according to both the user's own state and the summarized state of the neighbors.
Experiments on two public datasets show that the proposed model outperforms the state-of-the-art approaches significantly.

\end{abstract}

%
%


%

%
%

%
\maketitle

\section{Introduction}

With the development of the mobile internet, people now tend to read news online rather than reading conventional paper-based media. Such transition provides a great opportunity for personalized news recommendations.

Unlike E-commerce and many other scenes, active users can consume news very fast, for example, browsing over dozens of news articles within ten minutes. As a result, how to capture the user's rapidly changing interests and react to the user's latest behavior becomes one of the key challenges of news recommendations~\cite{OkuraTOT17}. To address these issues, sequential recommendations~\cite{fang2019deep}, which utilize the user behavior sequences and embed previously purchased products for current interest prediction~\cite{donkers2017sequential, tang2018personalized}, have been applied for news recommendations successfully, especially on some recent models based on Recurrent~Neural~Networks~(RNNs)~\cite{OkuraTOT17, KumarKG0V17}.

Despite the achievements, we argue that such methods still face challenges, mostly caused by the sociality in news reading. We can observe that users may read the news not because she was interested in the topic recently, but because she found it important~\cite{ozgobek2014survey}. People tend to follow these constantly changing ``hot'' topics, and sometimes the hot topics are only of significance to certain groups of people. With such sociality, predicting the relevance between users and news solely based on the target users' own browsing activities can be tough. Let's consider the following example:

\emph{Tom is a driver who likes baseball. One day, after he read a dozen baseball-related news articles, a piece of news entitled ``Ice is causing trouble in the traffic'' emerged and spread among people who were sensitive to the condition of traffic.\footnote{This example is inspired by the case studies in Section~\ref{sec:cases}}}

In this example, RNN-based sequential recommendations can hardly succeed in recommending the merging news. Some works try to alleviate this problem by considering the users' general preference~\cite{tang2018personalized, donkers2017sequential} or social relationships~\cite{song2019session}, however, we argue that using general preferences is too static, while the social relationships among users are hard to obtain. On the other side, traditional User-based~Collaborative~Filtering~(UserCF) approaches recommend news articles based on what other similar people (i.e., neighbors) are reading, which provide a possibility to jump out of the user's own browsing context and recommend the news that neighbors are interested in. However, the classic UserCF has some limitations, including: (1)~It's difficult for UserCF to react immediately to the target user's most recent behavior ~\cite{linden2003amazon}; (2)~UserCF only use one single scalar of similarity metric to represent the relationship between two users, which is too coarse-grained. As a result, some researchers suggest that UserCF is not suitable for news recommendations~\cite{zhong2015building, OkuraTOT17}, although the idea of using similar people's behaviors is valuable.

To tackle this challenge, in this paper, we integrate the RNN-based sequential recommendation algorithms and the key idea of User-based collaborative filtering into a framework of deep neural networks, and propose a new model called Collaborative~Sequential~Recommendation~Networks~(CSRNs). The proposed model works in the following way. 

Firstly, to better model the relationships between users, we design a directed news co-reading network, which is built with early browsing history. With Singular~Value~Decomposition~(SVD), users can be assigned with vectors indicating their general preferences, and similar users can be detected and modeled in a co-reading network. Furthermore, the edges between users can be described with a function of the vectors of connected users. Therefore, compared with UserCF, the relationships between users can be described in a vector space, which is a more fine-grained way.

Secondly, recommendations are made according to the recent browsing history of both the target user and her neighbors. Similar to other sequential recommendation models, we use RNNs to encode users' recent browsing history, thus the encoded features can represent what kind of news the users are interested in recently. Afterward, based on the current states of the target user and the neighbors, as well as their relationships, the model will pay different attention dynamically to different neighbors and make personalized summarization of what the neighbors are reading. Finally, recommendations are made according to the information from both the target user and the neighbors aspects.

In this way, the model can successfully handle the situation in the example by finding

\emph{Jerry and Tom share similar interests in traffic-related news. One day Jerry reads a piece of news in that category, and even if Tom didn't show any evidence of relevance in his recent behavior, the news could still be recommended to Tom.}

To the best of our knowledge, this is the first attempt to integrate the idea of UserCF into a neural network model for sequential recommendations. The key contributions of this paper are summarized as follows:
\begin{itemize}
    \item We propose an approach to build news co-reading networks, which can describe the relationships meticulously. The co-reading network will be employed for better news recommendation.
    \item Based on the news co-reading network, we propose a novel neural network model named as Collaborative~Sequential~Recommendation~Networks~(CSRNs), which combines the advantages of RNN-based sequential recommendations and UserCF.
    \item We evaluate the proposed model comprehensively on two public datasets. Extensive experimental results show that the proposed model outperforms the state-of-the-art baselines significantly.
\end{itemize}

The rest of this paper is organized as follows. We briefly summarize related works in Section~\ref{sec:related_work}. Section~\ref{sec:co-reading_network} introduces the design of the news co-reading network. The details of the proposed CSRNs are then presented in Section~\ref{sec:cfn}. Section~\ref{sec:escpn} conducts empirical studies on the co-reading network built with a real-world dataset. Section~\ref{sec:experiments} reports the experimental results, and Section~\ref{sec:conclusion} concludes the paper.

\section{Related Work}
\label{sec:related_work}

In this section, we review the literature on two topics that are most relevant to our research, i.e., news recommendations and sequential recommendations. We also explain how our approach differs from the literature.

\subsection{News Recommendations}

News recommendations have been widely studied in the research community. Early work used memory-based and model-based collaborative filtering algorithms~\cite{das2007google}. However, since news expires fast, CF-based methods often suffer from the cold-start problem. Therefore, many algorithms that utilized the content of news were proposed~\cite{bansal2015content, lu2015content}. For example, Lu et al.~\cite{lu2015content} proposed a content-based collaborative filtering approach to bring both content-based filtering and collaborative filtering approaches together. Recently, neural network-based algorithms have been widely studied for news recommendations~\cite{OkuraTOT17,KumarKG0V17, park2017deep}. Okura et al.~\cite{OkuraTOT17} used Recurrent Neural Networks~(RNNs) with users' recent browsing history to model user preference and make news recommendations. Kumar et al.~\cite{KumarKG0V17} proposed to use Bi-directional LSTM and self-attention to improve the accuracy of recommendations. Park et al.~\cite{park2017deep} used a Convolutional Neural Network model to capture user preferences and to personalize recommendation results. There are also many works that combine other features into news recommendations, including knowledge graph~\cite{wang2018dkn}, location~\cite{son2013location} and so on. Karimi et al. reviewed the state-of-the-art of designing and evaluating news recommender systems over the recent years in~\cite{karimi2018news}.

The major difference between prior works and ours is that we highlight the sociality in news reading by integrating the idea of UserCF with conventional RNN-based sequential recommendations. By building news co-reading network and utilizing the information from neighbors, better recommendations can be achieved.

\subsection{Sequential Recommendations}

Traditional collaborative filtering models often ignore the temporal information~\cite{su2009survey}. However, in the real world, the order of historical behaviors matters a lot. To model this phenomenon, many sequential recommendation algorithms have been proposed. Rendle et al.~\cite{rendle2010factorizing} applied Markov chains to model user behavior sequences, and later neural network-based methods have been proposed and significantly improved the performance~\cite{donkers2017sequential,tang2018personalized,chen2018sequential}. For example, Donkers et al.~\cite{donkers2017sequential} proposed a new type of Gated Recurrent Units incorporating the explicit notion of the user for whom recommendations are specifically generated. Tang et al.~\cite{tang2018personalized} proposed a convolutional sequence embedding recommendation model to address the union-level sequential patterns and skip behaviors. Although focused on different aspects, many RNN-based session-based recommendation algorithms actually share related model structures with sequential recommendation models~\cite{HidasiKBT15, HidasiK17}. The difference between session-based recommendation and sequential recommendation is that, users are often assumed to be anonymous under session-based recommendations, while for sequential recommendations, user IDs can be obtained to link the different sessions of a user together.

While existing sequential recommendation algorithms are trying to utilize the behaviors of target users better, to the best of our knowledge, the proposed CSRN is the first attempt to bring the information from the neighbors into sequential recommendations in a Neural Network model. Jannach et ta. found that combining the kNN approach with GRU4Rec can give better results for session-based recommendations in~\cite{jannach2017recurrent}, however, they only tried weighted averaging the results given by kNN and GRU4Rec, which is completely different from our work.




\section{Co-reading Network Construction}
\label{sec:co-reading_network}

To identify users with similar interest and describe such relationships, we introduce the news co-reading network, which is built with users' early browsing history. Let
\begin{displaymath}
  r_{ij} = \left\{
  \begin{array}{ll}
    1, & \text{user $i$ has interactions with news $j$}\\
    0, & \text{otherwise}
  \end{array} \right.
\end{displaymath}
indicate whether user $i$ has read the news $j$, and
\begin{displaymath}
  \mathbf{R} = [r_{ij}]_{i=1,j=1}^{I,J}
\end{displaymath}
is the $I \times J$ binary rating matrix of the early browsing history, $I$ is the number of distinct users and $J$ is the number of distinct news. We show how the news co-reading network is built with $\mathbf{R}$.

Firstly, we apply Singular~Value~Decomposition~(SVD) upon the TF-IDF transformed matrix $\mathbf{R}$, and only keep the $T$ largest singular values and the corresponding vectors, i.e.,
\begin{equation}
\label{eq:svd}
\hat{\mathbf{R}} \approx \mathbf{U}_{I \times T} \mathbf{\Sigma}_{T \times T} \mathbf{V}_{T \times J} \text{,}
\end{equation}
where $\hat{\mathbf{R}} = \text{TF-IDF}(\mathbf{R})$, and the TF-IDF transformation here is applied to down-weight the most popular news items.

$\mathbf{U}$ is an $I \times T$ matrix, and the $i$th row of $\mathbf{U}$, i.e., $\bm{u}_i$, can be a representation of user $i$. We consider user $i$ and user $k$ are neighbors if the similarity of embeddings satisfies some certain conditions. For example, we can keep the top $N$ most similar users as neighbors just like UserCF. The similarity scores are defined as:
\begin{equation}
\label{eq:user_sim}
\text{similarity}_{ik} = \bm{u}_{i} \mathbf{\Sigma} \bm{u}_{k}^{\text{T}}
\end{equation}

Secondly, to represent the relationship between user $i$ and one of her neighbors, \emph{i.e.}, user $k$, inspired by~\cite{gong2018adaptive}, we use a directed edge $\mathbf{e}_{ik}$ with features
\begin{equation}
  \bm{e}_{ik} = [\bm{u}_{i} \quad \bm{u}_{k} \quad \bm{u}_{i} \odot \bm{u}_{k}] \text{,}
\end{equation}
where $\odot$ is the \emph{Hadamard} product operator which computes element-wise multiplication between vectors. The three parts of $\bm{e}_{ik}$ describe the target user, the source user, and their undirected relationship respectively. Compared with traditional user-based collaborative filtering, $\bm{e}_{ik}$ describes the relationship in a more meticulous way.

Finally, we can define the news co-reading network with a set of triplets
\begin{displaymath}
  G=\{\langle (i, k), \bm{e}_{ik} \rangle \} \text{.}
\end{displaymath}
With the network, we can make better recommendations by CSRN.

Note that the network of users can also be built with information other than early browsing histories. For example, we can build a co-purchasing network using APP purchasing records. The intuition is that, users have similar APPs installed might be interested in similar kind of news~\cite{liu2017transfer}. By building a network with information other than the news reading domain, we can easily transfer knowledge and potentially handle the user-side cold-start problem.

\section{Collaborative Sequential Recommendation Networks}
\label{sec:cfn}

\begin{figure*}[!t]
\includegraphics[width=13cm]{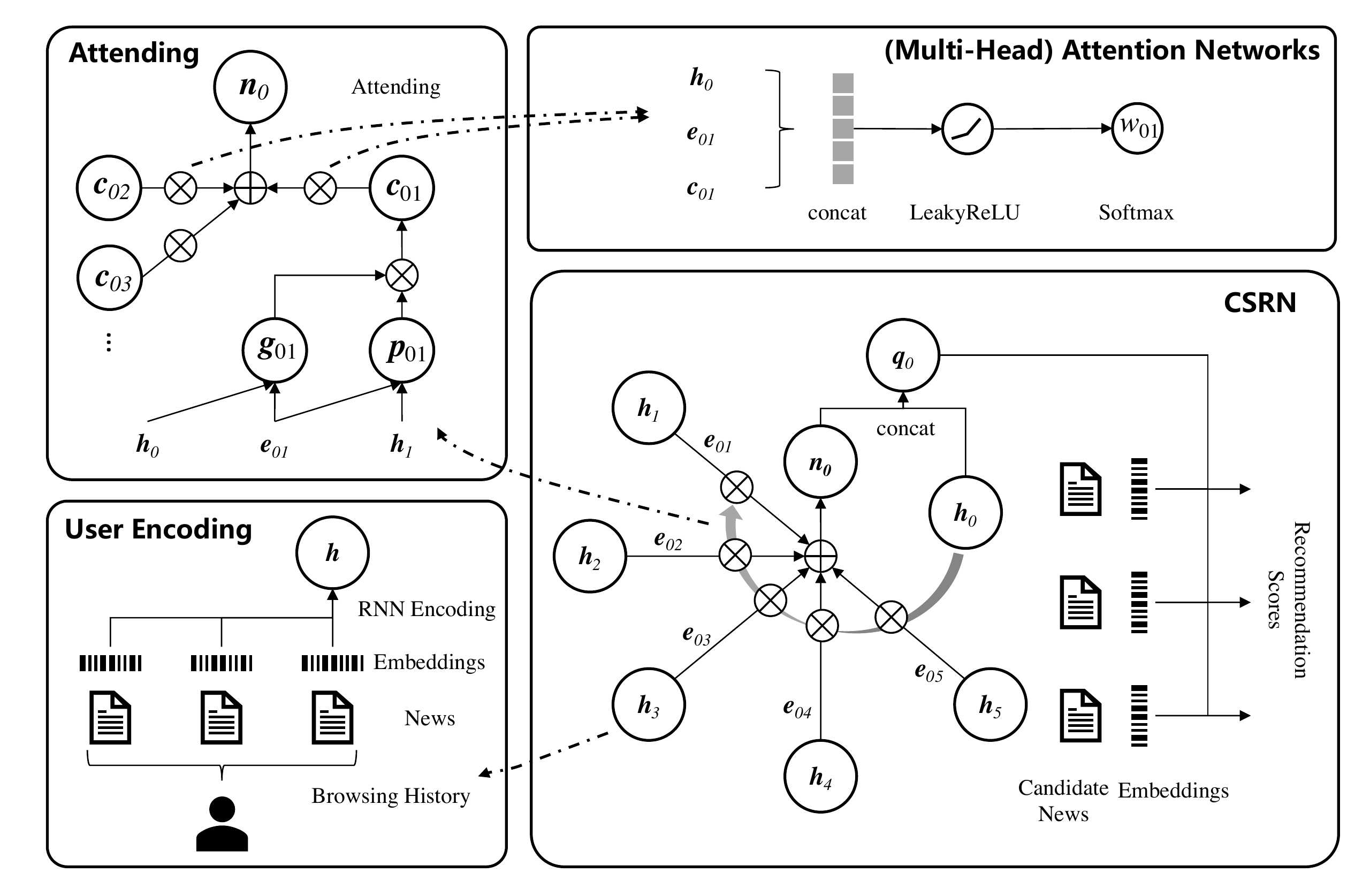}
\caption{The framework of the proposed CSRN. For demonstration, we assume that user 0 is the target user, and user 1--5 are the neighbors. Users are encoded with their recent browsing history, and based on the co-reading network, the model will learn to summarize what the neighbors are reading at the moment. Final recommendations are made with both $\bm{h}_0$ and $\bm{n}_0$.}
\label{fig:framework}
\end{figure*}

In this section, we explain the details of the proposed CSRN, which utilize the co-reading network of users to make news recommendations. The framework is illustrated in Figure~\ref{fig:framework}. The model consists of mainly three parts, i.e., user encoding, attending, and recommending.

\subsection{User Encoding}

We use RNNs to encode users' recent behaviors similar to~\cite{OkuraTOT17, HidasiKBT15}. RNNs are networks that deal with sequential data, and they compute the user's current hidden state based on the hidden state at last step and the input at the current step, i.e.,
\begin{displaymath}
  \bm{h}^t = f(\bm{v}^t, \bm{h}^{t-1}) \text{,}
\end{displaymath}
where $\bm{h}^t$ is the hidden state at step $t$, and $\bm{v}^t$ is the input vector at step $t$. There are many formulation of $f$, including Long~Short-Term~Memory~(LSTM)~\cite{hochreiter1997long}, Gated~Recurrent~Unit~(GRU)~\cite{cho2014properties}, vanilla RNN, and many other variants. 

For news recommendations, the inputs are the sequences of news embeddings that the users read recently, and the outputs are the hidden states of users. Details can be found in~\cite{OkuraTOT17, KumarKG0V17}. Note that a variety of user encoding models are compatible with the framework of CSRNs, as long as that the model can map a user to a vector.

\subsection{Attending}

Based on the co-reading network $G$, CSRN can learn to personally summarize what one's neighbors are interested in at the moment, and utilize the information to make better recommendations. As shown in Figure~\ref{fig:framework}, assuming that user $i$ is the target user to recommend for, and user $k$ is one of user $i$'s neighbors in the co-reading network $G$, $\bm{h}_i$ and $\bm{h}_k$ are users' hidden states encoded according to what news has been read recently, and $\bm{e}_{ik}$ is the feature of the directed edge connecting user $i$ and user $k$. The attending procedure works in the following way.

The first step is to decide what information can go through the edge. We use a single layer network defined by the following equation:
\begin{equation}
\label{eq:input}
    \bm{p}_{ik} = \phi (\mathbf{W}_{ph}\bm{h}_k + \mathbf{W}_{pe}\bm{e}_{ik} + \bm{b}_{p}) \text{,}
\end{equation}
where $\phi(\cdot)$ is the \emph{Tanh} function, and $\bm{p}_{ik}$ contains the information that can pass through the edge from user $k$ to user $i$.

The next step is to decide what information user $i$ would like to extract from the attended user $k$ based on the current state. This is achieved by the gates~\cite{cho2014properties} defined with the following equation:
\begin{equation}
\label{eq:gate}
  \bm{g}_{ik} = \sigma (\mathbf{W}_{gh}\bm{h}_i + \mathbf{W}_{ge}\bm{e}_{ik} + \bm{b}_{g}) \text{,}
\end{equation}
where $\sigma(\cdot)$ is the \emph{Sigmoid} function, and $\bm{g}_{ik}$ is the gate indicating what information user $i$ would like to extract from user $k$.

Then, the encoded information from user $k$ to $i$ is formulated as:
\begin{equation}
  \bm{c}_{ik} = \bm{g}_{ik} \odot \bm{p}_{ik} \text{.}
\end{equation}

The final step is to summarize the information from all the neighbors. Inspired by~\cite{Velickovic17}, we use (multi-head) attention networks to compute the weights for different neighbors. The attention networks work as follows:
\begin{equation}
  \alpha_{ik} = \text{LeakyReLU}(\bm{w}_{ah}\bm{h}_i + \bm{w}_{ae}\bm{e}_{ik} + \bm{w}_{ac}\bm{c}_{ik} + b_a)
\end{equation}
where the negative input slope of \emph{LeakyReLU} is set to 0.2. Then a \emph{Softmax} function is used to normalize the weights:
\begin{equation}
  w_{ik} = \frac{\text{exp}(\alpha_{ik})}{\sum_{l \in \mathbb{N}_i}\text{exp}(\alpha_{il})} \text{,}
\end{equation}
where $\mathbb{N}_i$ is the set of neighbors of user $i$.

The final summarization of the information from neighbors are defined as:
\begin{equation}
\label{eq:attending}
  \bm{n}_i = \sum_{k \in \mathbb{N}_i} w_{ik}\bm{c}_{ik}
\end{equation}

To stabilize the learning process of attention, we apply a multi-head attention mechanism. $H$ independent attention procedures executing the averaging transformation of Equation~(\ref{eq:attending}) are conducted, and the outputs are concatenated as the final summarization of the neighbors, i.e.,
\begin{equation}
  \bm{n}_i = \Big\|_{h=0}^{H}(\sum_{k \in \mathbb{N}_i} w_{ik}^h \bm{c}_{ik}^h) \text{,}
\end{equation}
where $\|$ represents the concatenation operation.

From the steps above, we can see how the features of edges which represent the relationship between users, and the hidden states of users which indicate their recent interests, are taking effects for news recommendations. The $\bm{n}_i$ contains personalized summarization of what the neighbors are reading, and can be valuable for better recommendations.

\begin{figure*}[!t]
\centering
\subfloat[Distribution of the out-degrees of nodes and the corresponding user activities. For demonstration purposes, only the nodes with out-degrees greater than 0 are included in this figure.]{\includegraphics[height=1.12in]{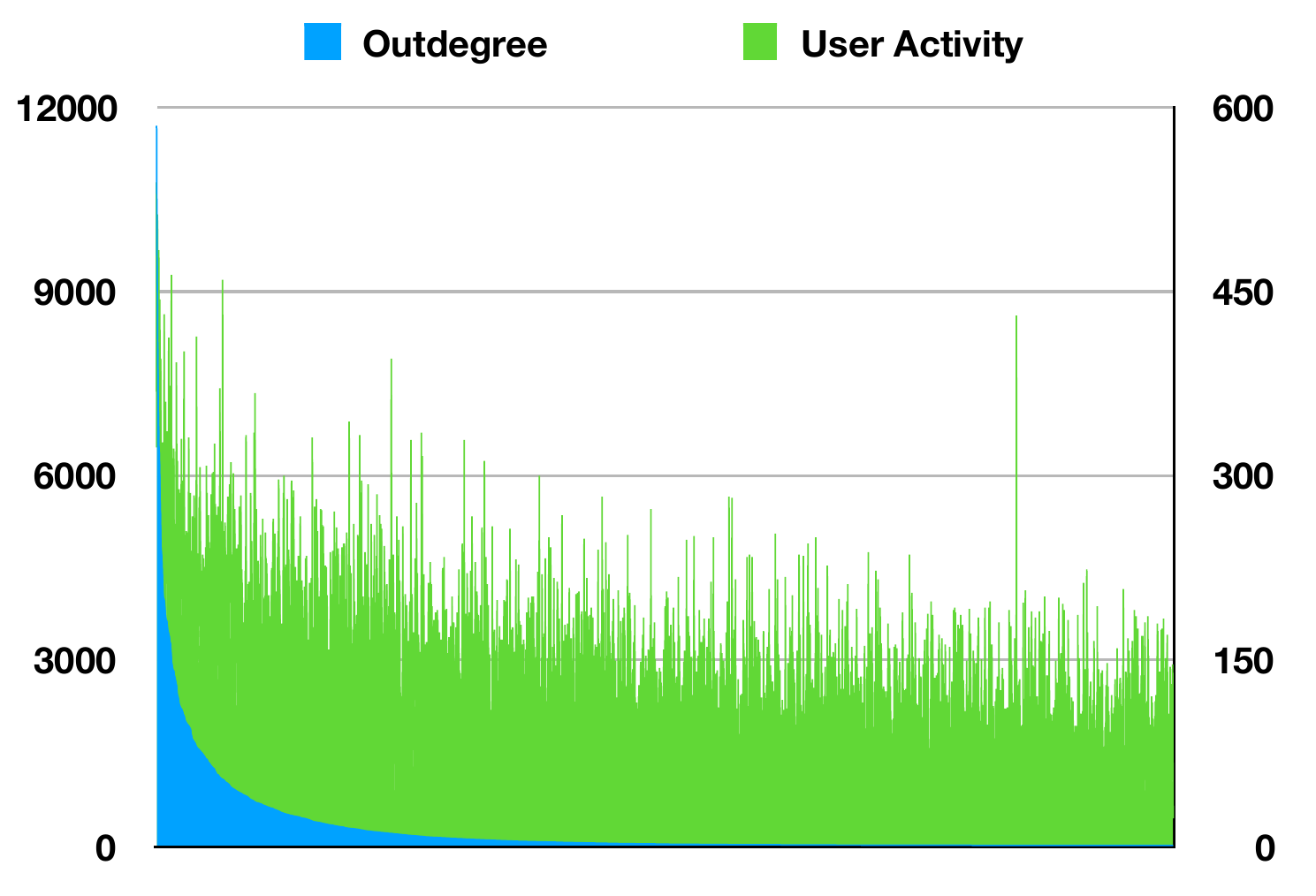}
\label{fig:cpn}}
\hfil
\subfloat[Percentage of node pairs that it can be reached within certain numbers of steps from a start node with out-degree to an end node.]{\includegraphics[height=1.12in]{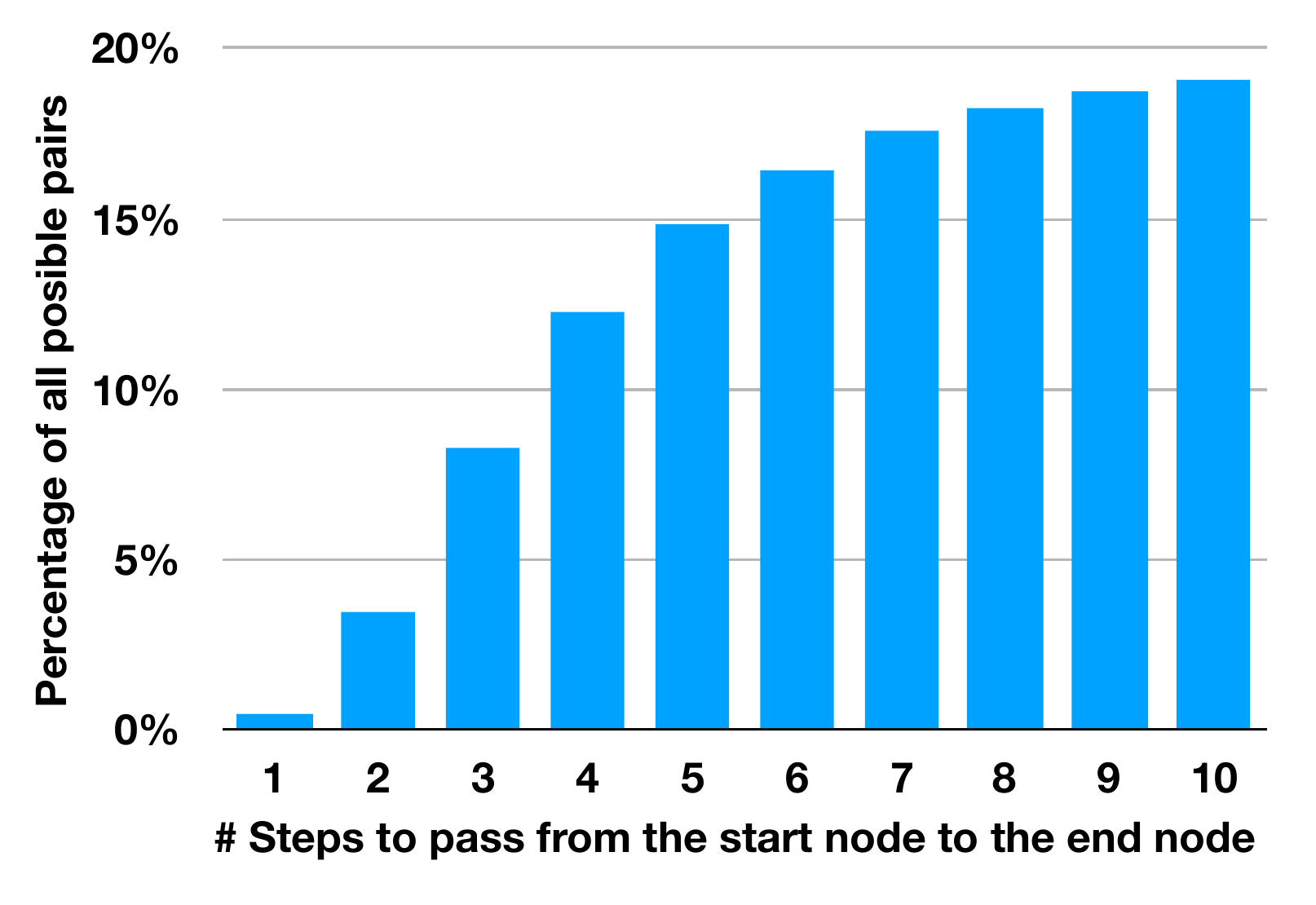}
\label{fig:pass_in_steps}}
\hfil
\subfloat[Visualization of the adjacency matrix of the top 1000 most active users with t-SNE. Each symbol in the figure represents a user, and the shapes of symbols represent the most clicked news category by the users.]{\includegraphics[height=1.12in]{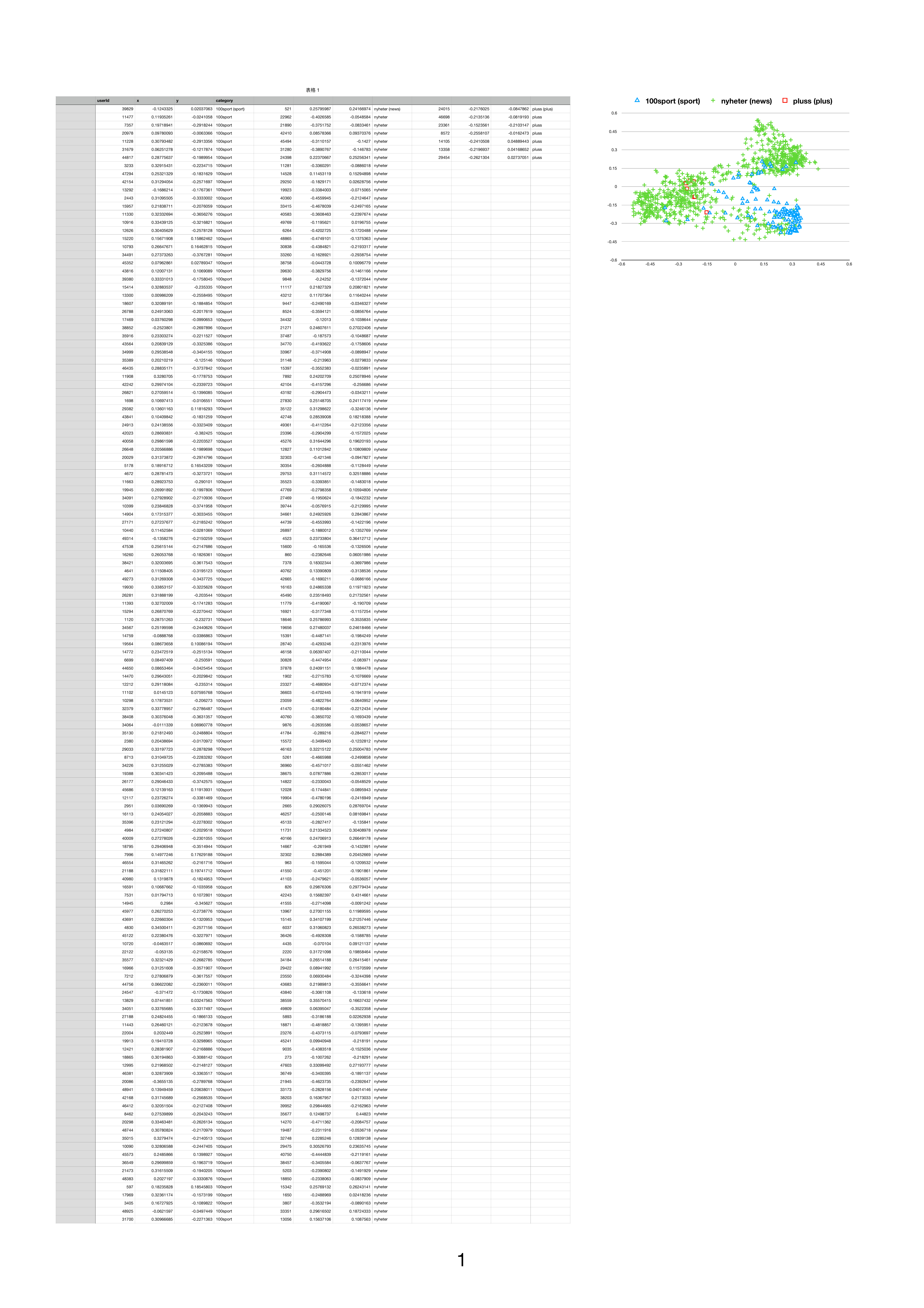}
\label{fig:user_vis}}
\hfil
\subfloat[Statistics about the proportions of clicks versus how many neighbors had already read the news before the clicking event.]{\includegraphics[height=1.12in]{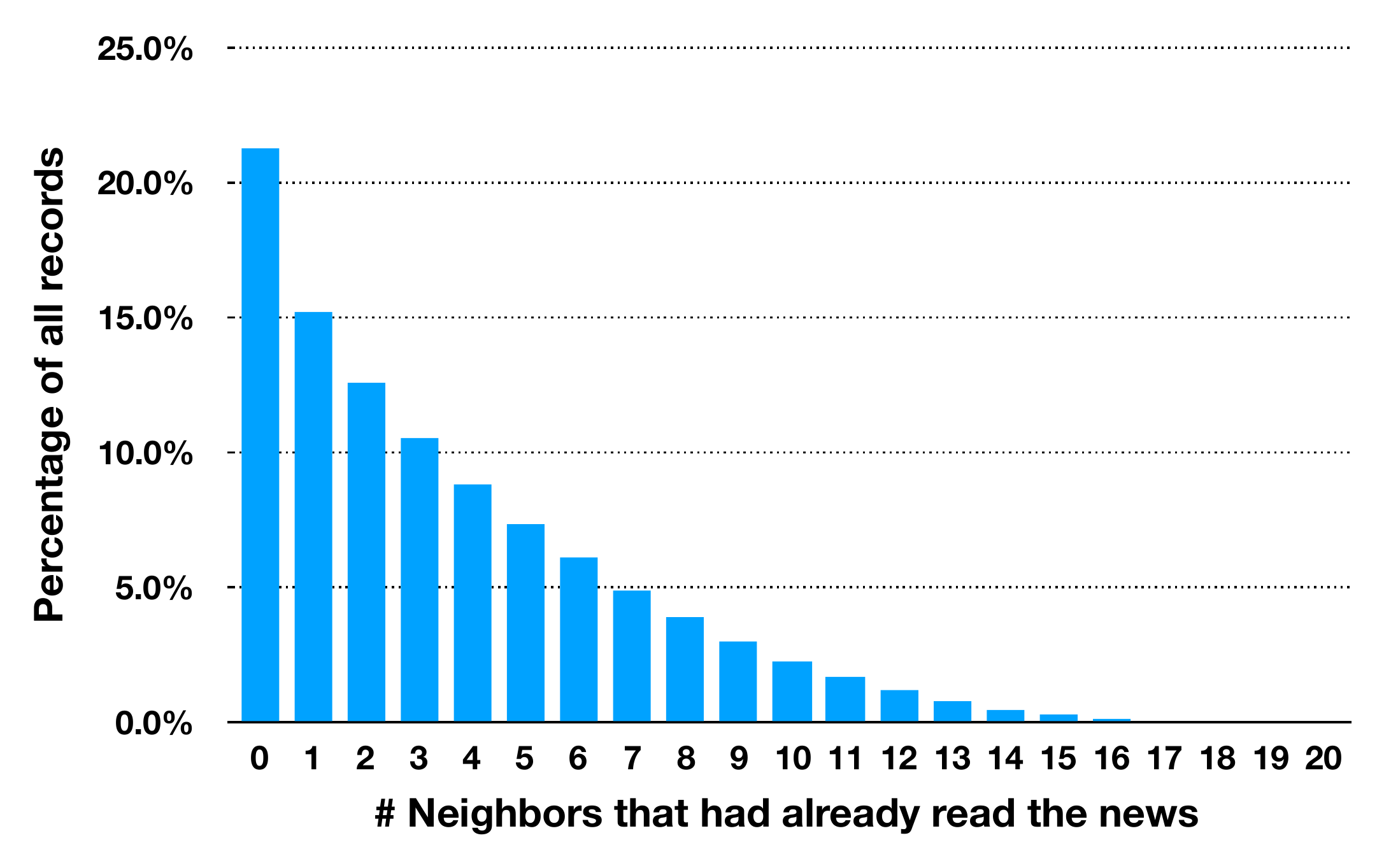}
\label{fig:click_times}}
\caption{Empirical studies on the co-reading network. Best viewed in color.}
\label{fig:examples}
\end{figure*}

\subsection{Recommending}

With the hidden states of the target user, i.e., $\bm{h}_i$ and the summarization of what the neighbors are reading, i.e., $\bm{n}_i$, the model can finally compute recommendation scores for each candidate news. In this paper, we deal with news retrieval tasks, thus the relevance function is restricted to a simple inner-product function~\cite{OkuraTOT17}.

The decoding function for users are formulated as:
\begin{equation}
\label{eq:decoding}
  \bm{q}_i = \phi(\mathbf{W}_{qh}\bm{h}_i + \mathbf{W}_{qn}\bm{n}_i + \mathbf{b}_q)
\end{equation}
and the recommendation score is defined as:
\begin{equation}
  r_{ij} = \bm{q}_i^\text{T} \bm{v_j} \text{,}
\end{equation}
where $v_j$ is the embeddings of news $j$.

\subsection{Loss Functions and Regularizations}

We experiment three kinds of loss functions, including two pair-wise losses \emph{TOP1-Max} and \emph{BPR-Max} proposed in~\cite{HidasiK17}, as well as the classic \emph{Cross-Entropy}. The definitions of the loss functions are illustrated as follows:
\begin{align}
    \label{eq:top1-max} & \mathscr{L}_{top1\text{-}max} = \sum_{j} s_j\big(\sigma(r_j - r) + \sigma(r_j^2)\big) \\
    \label{eq:bpr-max} & \mathscr{L}_{bpr\text{-}max} = -\text{log} \sum_{j} s_j \sigma(r - r_j) + \sum_{j} s_jr_j^2 \\
    \label{eq:xe} & \mathscr{L}_{xe} = -\text{log} \frac{\text{exp} (r)} {\text{exp} (r) + \sum_{j} \text{exp} (r_j)} \text{,}
\end{align}
where $r$ is the predicted rating for the positive sample, $r_j$ is the predicted rating for a negative sample, and $s_j$ is the weight for $r_j$, defined as:
\begin{equation}
\label{eq:sj}
  s_j = \frac{\text{exp}(r_j)}{\sum_{l \in \mathbb{N}_s} \text{exp} (r_l)} \text{,}
\end{equation}
where $\mathbb{N}_s$ is the set of negative samples. Detailed explanations about the loss functions can be found in~\cite{HidasiK17}.

Apart from the score regularization~\cite{HidasiK17} in TOP1-Max and BPR-Max, two kinds of model regularization are used, including weight decay and dropout~\cite{srivastava2014dropout}. Weight decay is applied to all the learnable parameters, and two dropout layers are used in the model. Firstly, the news embeddings are randomly masked off before being fed into the recurrent neural networks. Secondly, dropout is applied for the input of Equation~(\ref{eq:decoding}), i.e., before the final decoding function.

\section{Empirical studies on the co-reading network}
\label{sec:escpn}





To gain a better insight into the co-reading network, in this section, we conduct empirical studies on the co-reading network built with the Adressa dataset\footnote{Details about the dataset is introduced in Section~\ref{sec:datasets}}. We keep the default of top 20 most similar users as the neighbors for each user, unless other specific configurations are explicitly specified. As a result, every node in the network has an in-degree of 20, while the out-degrees can vary a lot.

Figure~\ref{fig:cpn} demonstrates the distribution of out-degrees of nodes (i.e., users). The out-degree shows a very skewed pattern, even more skewed compared with the distribution of user activities, which is defined by how many news articles each user read in the training set. Based on our experiments, only 3,776 out of 50,000 users in the dataset have out-degrees greater than 0, and the maximum out-degree is 11,680. We also find that the user with the largest out-degree clicked 538 news in the training set, which is not a significantly larger number compared with other active users. Although the out-degree is correlated with the user activity, the relationship is not absolute. Our conclusion is that the TF-IDF transformation in Equation~(\ref{eq:svd}) is taking effect, since if a user shows no significant preference on certain kinds of news and only reads the most popular ones, the value from her browsing history will be little. 

Figure~\ref{fig:pass_in_steps} focuses on the connectivity of the co-reading network. Starting from nodes with out-degrees greater than 0, we calculate the percentage of node pairs that it can be reached within certain steps out of all possible node pairs. As illustrated, only less than 20\% node pairs can be reached within 10 steps, and the percentage starts to saturate. This shows that the connectivity of the network is not strong, indicating that the interests of the users are significantly different, as the user aggregates into clusters that are separated from each other in the network. We further visualize the top 1,000 most active users with t-SNE~\cite{maaten2008visualizing} using the adjacency matrix as inputs in Figure~\ref{fig:user_vis}. The shapes of symbols represent the most clicked news category by the user. As we can see, there are three major clusters among them, and although all labeled as ``nyheter (news)'', users can still be divided into two groups. This phenomenon suggests that there is a large space for personalization, since the interests of users differ from each other significantly.

In order to explore the potential of utilizing neighbors to make news recommendations, we calculate the proportions of news clicking events in the training set versus how many neighbors had already read the news before the clicking event took place. Statistics are shown in Figure~\ref{fig:click_times}. We can find that 78.7\% of all the click events happened after at least one neighbor had read the clicked news already, and about 31.7\% of the click events took place when the clicked news had been read by equal to or more than 5 neighbors.

As a conclusion, taking the Adressa dataset as an example, we find that the co-reading network is highly centralized and poorly connected. This is a preferred property since it indicates that we only need to focus on a small set of representative users as neighbors for others. These users are usually very active, and have their own preferences for the type of news. Their browsing actions can be valuable for revealing the hot topics at early stages, and help to recommend news for other people better. Moreover, it's promising to utilize the browsing histories of neighbors, since a large share of browsing events happened after that the clicked news had already been viewed by multiple neighbors. What the models need to do is finding the shared interests among the people, and recommend corresponding news to the target users.

\section{Experimental Results}
\label{sec:experiments}

In this section, we discuss the experimental results. The datasets used for evaluation and the baselines are introduced first, followed by the evaluation scheme and how the hyper-parameters are decided. Evaluation results and case studies are presented at last.

\subsection{Datasets}
\label{sec:datasets}

\begin{table}[!t]
\renewcommand{\arraystretch}{1.0}
\caption{Information about the datasets}
\label{tb:dataset}
\centering
\begin{tabular}{p{3.5cm} p{2cm} p{2cm}}
\hline
Item & Adressa dataset & plista dataset \\
\hline
\# users & 50,000 & 30,000\\
\# news & 17,453 & 2,146\\
\# clicks in training set & 2,285,316 & 2,726,891\\
Duration of training set & 31 days & 12 days \\
\# clicks in validation set & 344,407 & 368,214\\
\# clicks in testing set & 344,408 & 368,215\\
Duration of testing set & 9 days & 6 days \\
Avg. \# words per news & 122.80 & 28.18\\
Density & 0.34\% & 5.38\% \\
\hline
\end{tabular}
\end{table}

We use two publicly available datasets for evaluation, including \emph{the Adressa dataset}~\cite{gulla2017adressa} and \emph{the plista dataset}~\cite{kille2013plista}. We use the user-news interaction relationships as well as the content of news. Other attributes of users are beyond the scope of this paper.

We extract the sequences in which the news articles were read by users for both datasets, and split the datasets into history / training / validation / testing sets. The history sets are used for constructing co-reading networks. Detailed steps of dataset preprocessing are introduced in Appendix~\ref{sec:dataset_preprocessing_detailed}. Statistics about the datasets are listed in Table~\ref{tb:dataset}.

\begin{table*}[!t]
\renewcommand{\multirowsetup}{\centering}
\renewcommand{\arraystretch}{1.0}
\caption{Evaluation results of the proposed CSRN and baselines}
\label{tb:evluation}
\resizebox{0.9\textwidth}{!}{
\begin{tabular}{c|p{1.25cm}<{\centering}|p{1.25cm}<{\centering}|p{1.25cm}<{\centering}|p{1.25cm}<{\centering}|p{1.25cm}<{\centering}|p{1.25cm}<{\centering}|p{1.25cm}<{\centering}|p{1.25cm}<{\centering}}
\hline
\multirow{2}*{\textbf{model}} & \multicolumn{4}{c|}{\textbf{Adressa dataset}} & \multicolumn{4}{c}{\textbf{plista dataset}}  \\
\cline{2-9}
 & \textbf{HR@1} & \textbf{HR@10} & \textbf{HR@20} & \textbf{MRR} & \textbf{HR@1} & \textbf{HR@10} & \textbf{HR@20} & \textbf{MRR} \\
\hline
POP & 0.61\% & 10.24\% & 20.48\% & 0.0482 & 0.02\% & 1.29\% & 6.98\% & 0.0239 \\
\hline
ItemCF & 1.29\% & 12.19\% & 23.66\% & 0.0600 & 1.56\% & 14.96\% & 28.81\% & 0.0704 \\
\hline
UserCF & 0.70\% & 10.07\% & 20.08\% & 0.0492 & 0.70\% & 10.08\% & 20.11\% & 0.0492 \\
\hline
BPR & 3.29\% & 24.00\% & 42.05\% & 0.1057 & 1.88\% & 14.55\% & 25.80\% & 0.0700 \\
\hline
GRU$_{top1\text{-}max}$ & 4.82\% & 33.25\% & 52.31\% & 0.1387 & \textbf{4.13\%} & 20.12\% & 35.25\% & 0.1038 \\
GRU$_{bpr\text{-}max}$ & 4.82\% & 33.20\% & 51.74\% & 0.1385 & 3.99\% & 20.80\% & 36.02\% & 0.1032 \\
GRU$_{xe}$ & 4.56\% & 33.73\% & 53.79\% & 0.1375 & 2.91\% & 19.68\% & 36.79\% & 0.0949 \\
\hline
Caser$_{top1\text{-}max}$ & 5.08\% & 33.78\% & 52.71\% & 0.1424 & 4.04\% & 20.36\% & 36.45\% & 0.1027 \\
Caser$_{bpr\text{-}max}$ & 4.97\% & 33.66\% & 52.20\% & 0.1413 & 4.01\% & 20.43\% & 36.52\% & 0.1027 \\
Caser$_{xe}$ & 4.82\% & 34.44\% & 54.43\% & 0.1417 & 2.91\% & 19.66\% & \textbf{36.89\%} & 0.0952 \\
\hline
AUGRU$_{top1\text{-}max}$ & \textbf{5.25\%} & 35.14\% & 53.93\% & \textbf{0.1468} & 4.03\% & 20.48\% & 36.03\% & 0.1030 \\
AUGRU$_{bpr\text{-}max}$ & 5.18\% & 35.22\% & 53.94\% & 0.1463 & 3.99\% & \textbf{21.15\%} & 36.39\% & \textbf{0.1045} \\
AUGRU$_{xe}$ & 5.00\% & \textbf{35.36\%} & \textbf{55.18\%} & 0.1454 & 2.96\% & 19.80\% & 36.89\% & 0.0956 \\
\hline
\hline
CSRN$_{top1\text{-}max}$ & \textbf{5.97\%} & 38.10\% & 57.61\% & 0.1603 & 3.99\% & 23.04\% & 39.28\% & 0.1075 \\
CSRN$_{bpr\text{-}max}$ & 5.91\% & 38.28\% & 57.54\% & \textbf{0.1603} & \textbf{4.00\%} & \textbf{23.76\%} & \textbf{40.18\%} & \textbf{0.1097} \\
CSRN$_{xe}$ & 5.69\% & \textbf{38.55\%} & \textbf{58.74\%} & 0.1582 & 2.86\% & 17.96\% & 32.43\% & 0.0899 \\
\hline
\hline
CSRN vs GRU ($bpr\text{-}max$) & +22.50\% & +15.27\% & +11.21\% & +15.72\% & +0.07\% & +14.26\% & +11.53\% & +6.23\% \\
CSRN vs Caser ($bpr\text{-}max$) & +18.96\% & +13.70\% & +10.22\% & +13.50\% & -0.21\% & +16.31\% & +10.01\% & +6.73\% \\
CSRN vs AUGRU ($bpr\text{-}max$) & +14.02\% & +8.67\% & +6.67\% & +9.56\% & +0.29\% & +12.32\% & +10.39\% & +4.91\% \\
\hline
\end{tabular}
}
\end{table*}

\subsection{Baselines}

We compare our proposed model with the following baselines.
\begin{itemize}
    \item \textbf{POP}~\cite{HidasiKBT15}: News articles are ranked by their popularities, i.e., how many times the news has been clicked by others.
    \item \textbf{ItemCF}~\cite{su2009survey}: This is one of the classical neighborhood-based collaborative filtering methods. News articles similar to what the user has read will be recommended to the target user. Instead of calculating similarities between news based on user-news interactions, we use the cosine similarities of embeddings. Experiments show that this modification can give better results since it can handle the item-side cold-start problem.
    \item \textbf{UserCF}~\cite{su2009survey}: This is another form of the classical neighborhood-based collaborative filtering. News consumed by similar users will be recommended.
    \item \textbf{BPR}~\cite{RendleFGS09}: This is a commonly used matrix factorization method that optimizes a pair-wise ranking loss.
    \item \textbf{GRU}~\cite{OkuraTOT17}: GRU are used for news sequential recommendation in~\cite{OkuraTOT17}. To overcome the unavoidable cold-start problem of news, embeddings are first learnt from the content of news articles. We follow the improvement in the loss proposed in~\cite{HidasiK17} in order to get a stronger baseline.
    \item \textbf{Caser}~\cite{tang2018personalized}: Caser considers union-level sequential patterns and skip behaviors by modeling user past historical interactions with both hierarchical and vertical convolutional neural networks. It also considers the users' general preferences.
    \item \textbf{AUGRU}~\cite{donkers2017sequential}: Attentional User-based GRU~(AUGRU) considers individual users' general preference in addition to sequences of consumed items. Attention mechanism is applied to adaptively shift focus between user and item aspects.
\end{itemize}

Compared with the proposed CSRN, UserCF utilizes a related basic idea while fails to describe the relationship between users in a fine-grained way. GRU only consider the target user's own browsing history, thus it can serve as an ablation experiment. Caser and AUGRU are state-of-the-art methods for sequential recommendations that consider both user activities and general preferences.

\subsection{Evaluation Scheme and Hyper-parameter Settings}
 
We adopt the \emph{leave-one-out} evaluation method for news retrieval tasks similar to~\cite{lian2018towards, KumarKG0V17}. The algorithms are requested to rank the ground truth news with 99 negative samples, and the negative samples are fixed and shared by all methods for fairness. The performance is judged by Hit~Rate~(HR) and Mean~Reciprocal~Rank~(MRR). More detailed evaluation scheme is introduced in Appendix~\ref{sec:evaluation_detailed}.

The news embeddings are learnt with CDAE proposed in~\cite{okura2016article} and shared by all the methods that require representations for the content of news. The embedding size is set to 256.

We use grid searching to find the best hyper-parameters for the baselines and the proposed CSRN according to MRR on the validation set, and report the corresponding results on the testing set. For ItemCF, best performances are achieved when the number of neighbors is set to 350 on the Adressa dataset, and 300 on the plista dataset. For UserCF, the number of neighbors is set to 150 on the Adressa dataset, and 250 on the plista dataset. All the hidden sizes for BPR, GRU, Caser, AUGRU and the proposed CSRN are set to 128, and the RNN parts of GRU, AUGRU and CSRN are all single layer GRU units. We use 4 attention heads, and the dimension of $\bm{c}_{ik}$ for each attention head is set to 32 to keep the dimensions of $\bm{h}_i$ and $\bm{n}_i$ equal.

The detailed hyper-parameter settings and the training scheme for NN-based methods are reported in Appendix~\ref{sec:hypers_detailed}.

\subsection{Evaluation Results}
\label{sec:evaludation_results}

The evaluation results are reported in Table~\ref{tb:evluation}. We report HR@1, HR@10, HR@20, and MRR results of the algorithms on both datasets.

As we can see, methods that tend to bias towards popular items significantly, i.e., POP and UserCF, don't perform well on both datasets, partially because of that the negative sampling strategy is related to the popularities of items. The models need to distinguish between general popularity and personal relevance, thus this is a relatively hard setting especially for these methods. In our experiments, ItemCF outperforms UserCF, mainly because that we use the cosine similarities of news embeddings in ItemCF, so it can recommend cold-start news. Matrix factorization-based methods, i.e., BPR, get the best results among traditional recommendation algorithms.

On both datasets, GRU, Caser and AUGRU outperform traditional baselines significantly. Comparing the results of these algorithms, we find that Caser and AUGRU enjoy larger promotions on the Adressa dataset than on the plista dataset, our assumption is that it is related to the density of the datasets. The user behavior on the plista dataset is far more dense than it on the Adressa dataset, the recent browsing histories are less likely to get outdated, thus introducing the users' general preferences cannot help very much.

Among all the tested methods, CSRN gets the best results under most of the metrics. Interestingly, CSRN enjoys a larger improvement under HR@1 and MRR, which are both sensitive to the accuracy of the very top part, on the Adressa dataset than on the plista dataset. Our conclusion is that this phenomenon is related to the densities of user behaviors too. The user's recent behavior on the Adressa dataset is more likely to expire, thus the information from the neighbors can help a lot for identifying the right news and put it to the very top.

Another interesting finding is that the TOP1-Max and BPR-Max loss functions show significantly superior performance than CrossEntropy loss on the plista dataset. We find that it is related to the score regularization. If the score regularization is disabled, we can find a phenomenon similar to the CrossEntropy loss, i.e., serious overfitting occurs and it cannot be settled by dropout and weight decay. This finding shows that the score regularization is beneficial to the learning process, as it can prevent the model from simply memorizing the negative samples, especially on the plista dataset which involves a limited number of distinct news and a relatively short duration of time.

\begin{table}[!t]
\renewcommand{\arraystretch}{1.0}
\caption{Impact of key hyper-parameters and ablation experiments on the Adressa dataset}
\label{tb:parameters}
\centering
\resizebox{0.485\textwidth}{!}{
\begin{tabular}{c | c c | c c}
\hline
 & Hyper-paramter & Value & HR@10 & MRR \\
\hline
base & & & 38.28\% & 0.1603 \\
\hline
\multirow{2}*{(A)} & \multirow{2}*{RNN Cell} & LSTM & 38.12\% (-) & 0.1591 (-) \\
 & & Vanilla RNN & 38.36\% (+) & 0.1602 (-) \\
\hline
\multirow{2}*{(B)} & Hidden Size & 512 & 38.06\% (-) & 0.1620 (+) \\
 & Hidden Layers & 2 & 38.17\% (-) & 0.1582 (-) \\
\hline
\multirow{3}*{(C)} & \multirow{3}*{\# Attention Heads} & 1 & 38.10\% (-) & 0.1600 (-) \\
 & & 2 & 38.23\% (-) & 0.1600 (-) \\
 & & 8 & 38.46\% (+) & 0.1602 (-) \\
\hline
\multirow{2}*{(D)} & \multirow{2}*{\# Neighbors} & 10 & 38.25\% (-) & 0.1588 (-) \\
 & & 30 & 38.39\% (+) & 0.1624 (+) \\
\hline
\multirow{4}*{(E)} & Edge Features & disabled & 37.48\% (-) & 0.1558 (-) \\
& Neighbor Selection & without TF-IDF & 38.03\% (-) & 0.1586 (-)\\
& Neighbor Selection & by random & 36.71\% (-) & 0.1477 (-) \\
& Neighbor Information & disabled & 33.20\% (-) & 0.1385 (-) \\
\hline
\end{tabular}
}
\end{table}

\begin{table*}[!t]
\renewcommand{\multirowsetup}{\centering}
\renewcommand{\arraystretch}{1.0}
\caption{Example cases and the ranking positions given by different algorithms}
\label{tb:case_study}
\noindent\makebox[\textwidth]{
\resizebox{0.9\textwidth}{!}{
\begin{tabular}{c | c | p{0.9cm}<{\centering} p{0.9cm}<{\centering} p{0.9cm}<{\centering} p{0.9cm}<{\centering} p{1.2cm}<{\centering}}
\hline
Case No. & Title of the Ground Truth News & GRU & Caser & AUGRU & CSRN & \# Viewed \\
\hline
\multirow{2}*{(1)} & Isen skaper trøbbel i trafikken & \multirow{2}*{57} & \multirow{2}*{41} & \multirow{2}*{34} & \multirow{2}*{3} & \multirow{2}*{13} \\
 & (Ice creates trouble in traffic) & & & \\
\hline
\multirow{2}*{(2)} & Ingen holdepunkt for å si at det er skutt mot bussen & \multirow{2}*{33} & \multirow{2}*{40} & \multirow{2}*{28} & \multirow{2}*{4} & \multirow{2}*{5} \\
 & (No clue to say it's shot on the bus) & & & \\
\hline
\multirow{2}*{(3)} & Norges tennissensasjon invitert til storturnering med stjernene & \multirow{2}*{21} & \multirow{2}*{9} & \multirow{2}*{5} & \multirow{2}*{3} & \multirow{2}*{13} \\
 & (Norway's tennis competition invited the stars to the tournament) & & & \\
\hline
\multirow{2}*{(4)} & Demidov til MLS & \multirow{2}*{37} & \multirow{2}*{35} & \multirow{2}*{44} & \multirow{2}*{4} & \multirow{2}*{5} \\
 & (Vadim Demidov transfered to MLS club) & & & \\
\hline
\multirow{2}*{(5)} & Offentlig rangering forsterker ulikhetene mellom skolene & \multirow{2}*{28} & \multirow{2}*{50} & \multirow{2}*{36} & \multirow{2}*{2} & \multirow{2}*{1} \\
 & (Public ranking reinforces inequalities between schools) & & & \\
\hline
\multirow{2}*{(6)} & Økte bomsatser er den mest effektive måten å redusere trafikken på & \multirow{2}*{18} & \multirow{2}*{10} & \multirow{2}*{19} & \multirow{2}*{5} & \multirow{2}*{0} \\
 & (Increased toll rates are the most effective way to reduce traffic) & & & \\
\hline
\end{tabular}
}
}
\end{table*}

\subsection{Impacts of Key Hyper-parameters and Ablation Experiments}


We study the impacts of some key hyper-parameters and some key components on the Adressa dataset and report the results in Table~\ref{tb:parameters}. We mainly focus on HR@10 which represents the performance of the recall ability, and MRR which focuses more on the order of the very top. We use the BPR-Max loss which performs well on both datasets.

Group A studies the impact of RNN cells. We find that while vanilla RNN can give comparable performance, LSTM tends to get overfitted a little. We briefly experimented LSTM with stronger regularization and find that using stronger dropout and weight decay cannot help significantly.

Group B studies the impact of model capacity. We find that simply adding the number of hidden units can lead to significantly better MRR results, while HR@10 decreases a bit. Adding more layers to the user encoding part cannot give better results under both metrics, confirming the conclusions from~\cite{HidasiKBT15}.

Group C studies the impact of multi-head attention. To prevent from adding too many parameters, we fix the size of $\bm{n}_i$, i.e., the representations after concatenation. Experiments show that multi-head attention can bring some improvements, especially for HR@10. However, too many heads may bring an adverse impact under MRR.

Group D studies the impact of the number of neighbors. Based on the results, we can see that adding more neighbors can significantly improve both metrics. However, it imposes more computation overhead. It's a trade-off between the computational cost and the performance.

Group E studies the impact of key components of CSRN. Firstly, we disable the edge features $\mathbf{e}_{ik}$ in Equation~(\ref{eq:input}) and (\ref{eq:gate}). Then, we try selecting neighbors without the TF-IDF transformation in Equation~(\ref{eq:svd}), and completely by random. Finally, we disable the information from the neighbors, \emph{i.e.}, $\mathbf{n}_i$ in Equation~(\ref{eq:decoding}), which gets the baseline GRU. Experiments show that all the components play a critical role in achieving the best performance.

\subsection{Case Studies}
\label{sec:cases}

To gain a better insight into the proposed CSRN, we take several cases from the Adressa dataset and study their effectiveness. Titles of the ground truth news, the ranks given by GRU, Caser, AUGRU as well as CSRN, and how many neighbors had already viewed the news are reported in Table~\ref{tb:case_study}. We use BPR-Max loss for case studies.

The news in Case~1 is about the impact of the bad weather. Intuitively, people might read dozens of news articles about sports, but they don't read that much news about the weather in a short term, so it's harder to infer the relevance of weather-related news from the user's browsing history, which is illustrated by the ranks given by other methods. However, this news is highly relevant to some specific groups of people. CSRN can find the relevance from the neighbors who share similar interest with the target user, thus put the news to the top successfully. Case~2 is about an attack at Saupstadringen, Norway, and we can find results similar to Case~1. By digging into what news the neighbors are reading, emerging news can be recommended accurately even if no evidence according to the user's own browsing history shows the intrinsic relevance. 

Case~3 and Case~4 are two examples of sport-related news, and the news of Case~3 is about a tennis competition, while the news of Case~4 is about a transfer event. Both news could be of great interests to certain groups of readers.  CSRN can successfully put the right news on the top of the recommendation list, thanks to the information from neighbors. These cases show that the proposed CSRN models can find the pattern that, if two users both like a certain category of news, and a news article of that category is viewed by one of them, then the news should be recommended to the other. In both cases, this strategy is more effective than simply using the users' static general preference.

Case~5 and Case~6 illustrate how the models would perform if only a few or none of the neighbors have read the ground truth article. As we can see from the table, even if no neighbors have viewed the news in Case~6, CSRN can still make better recommendations than the baselines sometimes. We dig into the data and find that in Case~6, neighbors were reading news articles entitled with ``48 000 flere trailere på veien hver dag (48,000 more trailers are on the road every day)'', ``Det er lov å sykle to i bredden (It is allowed to ride two in the width for cycling)'', and ``Mange busskur har blitt knust i Trondheim (Many buscars have been broken in Trondheim)''. CSRN can find that the neighbors were interested in traffic-related news, this information could spread along the co-reading network, and finally contributed to a better recommendation.

As illustrated by the cases above, what other similar users are reading can help a lot for news recommendations, and the proposed CSRN can find the pattern from data and give better results.

\section{Conclusion}
\label{sec:conclusion}

In this paper, we present the Collaborative Sequential Recommendation Networks, which integrate the RNN-based sequential recommendations with the idea of User-based collaborative filtering into the framework of deep neural networks. Firstly, we propose the methodology of building co-reading networks with users' early browsing history, then we propose the CSRN model which can learn attending functions of neighbors and make personal summarizations of what other users are reading. Using both the target user's recent browsing history and the summarization of what the neighbors are reading, better recommendations can be achieved. Comprehensive experiments on two publicly available datasets show that the proposed CSRN outperforms baselines significantly.

There are several potential extensions to CSRN that could be addressed in our future work. Firstly, explicit temporal information could help the model to decide when to trust the user's own browsing history more and when to trust information from the neighbors more. Secondly, we'd like to explore the performance of CSRNs on other domains, like movie or music recommendations. Finally, extending the model for cold-start users could be another interesting research direction.

%

%
\bibliographystyle{ACM-Reference-Format}
\bibliography{sample-base}

%
\appendix

\section{Experimental Setup}

\subsection{Dataset Preprocessing}
\label{sec:dataset_preprocessing_detailed}

For the Adressa dataset, we extract the sequences in which the news articles were read by users, and keep the data of the top 50,000 most active registered users. If a user read news for less than 5 seconds, the click is considered as a mis-action and discarded. Records before 2017-02-19 are treated as the history to construct the news co-reading network, records between 2017-02-20 and 2017-03-22 are used as the training set, and records after 2017-03-23 are then divided into a validation set and a testing set. For the plista dataset, we also extract the sequences in which the news articles were read by users, and keep the data of the top 30,000 most active users. Clicks on news without content are discarded since we cannot learn the embeddings for them. Records before 2016-02-10 are treated as the history, records between 2016-02-11 and 2016-02-22 are used as the training set, and records after 2016-02-23 are then divided into a validation set and a testing set. Since the news provided by the plista dataset is very limited, we crawl about 140k more news articles from www.tagesspiegel.de for news embedding learning. The categories of the crawled news are inferred from the URL.

\subsection{Detailed Evaluation Scheme}
\label{sec:evaluation_detailed}

We adopt the leave-one-out evaluation method similar to~\cite{lian2018towards, KumarKG0V17}. Given a user's most recent browsing history, only the next clicked news serves as the positive sample. Negative samples for training are dynamically drawn from a larger pool, and negative samples for validation and testing are drawn once and shared by all models to give a fair comparison.

The negative sampling pool is based on whether the news articles were clicked within a time interval, and only news that wasn't interacted by the users could be drawn as a negative sample. We find that the frequency of a news article being sampled as negative a sample is highly correlated to the popularity of that article. Since it's too time-consuming to rank all items, similar to~\cite{lian2018towards}, for validation and testing we draw 99 negative samples, which means the models need to rank among 100 news and find which one might be clicked by users. The performance is judged by Hit~Rate~(HR) and Mean~Reciprocal~Rank~(MRR) defined as follows,
\begin{displaymath}
\begin{split}
  \text{HR}@K & = \frac{1}{|\mathbb{C}|} \sum_{c \in \mathbb{C}} \mathbbm{1}(R_c \leq K) \\
  \text{MRR} & = \frac{1}{|\mathbb{C}|} \sum_{c \in \mathbb{C}} \frac{1}{R_c}
\end{split}
\end{displaymath}
where $\mathbb{C}$ is the set of clicks, $R_c$ is the ranking position of ground truth article for click event $c \in \mathbb{C}$, and $\mathbbm{1}(\cdot)$ is the indicative function.

While MRR is more sensitive to the accuracy of the very top part of the ranking list, HR@$K$ treats the top $K$ positions equally.

\subsection{Detailed Settings for Hyper-parameters}
\label{sec:hypers_detailed}

For co-reading news construction, we set $T$ to 32, i.e., we keep the 32 largest singular values and the corresponding vectors. We keep the default of top 20 most similar users as the neighbors for each user, unless other specific configurations are explicitly specified.

The news embeddings are learnt with CDAE and shared by all the baselines and CSRN which require representations of news articles. The embedding size is set to 256. The best hyper-paramters for CDAE is decided based on the MRR results given by GRU. For the Adressa dataset, we keep the 10,000 most frequent word tokens which appeared in less than 25\% articles as inputs, masking noise level is set to 0.3, and weight decay is set to 8e-5. Keywords and name entities provided by the dataset are concatenated with the documents. For the plista dataset, we keep the 25,000 most frequent word tokens which appeared in less than 20\% articles as inputs, masking noise level is set to 0.25, and weight decay is set to 1e-4. We find that the performance of algorithms is more sensitive to the hyper-parameters of CDAE on the plista dataset than it on the Adressa dataset, and the keywords and name entities in the Adressa dataset can help a lot for better embeddings.

Best hyper-parameters for baselines and CSRN are decided according to MRR on the validation set, and we report the corresponding results on the testing set. 

For neighborhood-based CF methods, the number of neighbors starts from 50 and increase 50 each time until the performance starts to decrease. For ItemCF, best performances are achieved when the number of neighbors is set to 350 on the Adressa dataset, and 300 on the plista dataset. For UserCF, the number of neighbors is set to 150 on the Adressa dataset, and 250 on the plista dataset.

All the hidden sizes for BPR, GRU, Caser, AUGRU and the proposed CSRN are set to 128. Weight decay is chosen from [1e-3, 1e-4, 1e-5, 1e-6], and we found that best weight decay for CSRN is 1e-5, while for other models it's 1e-4. Then we use grid searching to find the best hyper-parameters according to MRR on the validation set. The search range of dropout rate is [0.1, 0.15, 0.2, 0.25, 0.3, 0.35, 0.4]. For the Adressa dataset, dropout rate for GRU is set to 0.1 for the input embeddings and before the final decoder. For the plista dataset, dropout rate for GRU is set to 0.2 for the input embeddings and before the final decoder. For Caser, the max height $h$ is chosen from [1, 2, 4, 8], the number of horizontal filters for each height and the number of vertical filters are all chosen from [4, 8, 16, 32], and we found that best performance is achieved when they set to 4, 8, 16 respectively for the Adressa dataset, and 2, 8, 8 respectively for the plista dataset. For AUGRU, we find that the best dropout rates are 0.15 for input embeddings and 0.1 before the final decoder function on the Adressa dataset, and 0.25 for both on the plista dataset. For CSRN, we find that the best performance is achieved when the dropout rate is set to 0.15 for the input embeddings and 0.2 for the decoder with the Adressa dataset, and 0.2 for the input embeddings and before the decoder with plista dataset respectively. We use 4 attention heads, and the dimension of $\bm{c}_{ik}$ for each attention head is set to 32 to keep the dimensions of $\bm{h}_i$ and $\bm{n}_i$ equal.

For all the NN-based methods, we use the PyTorch\footnote{https://pytorch.org} for implementation with 2 NVIDIA Tesla M40 GPUs. RMSprop is used as the optimizer, batch size is set to 256. Learning rates start from 0.0001, and then decay at a fixed rate of every 1000 steps. Gradient clipping is used to avoid gradient explosion and set to 5. We use single layer GRUs for RNN-based methods if no specific configuration is mentioned. 

\end{document}